\documentclass[conference]{IEEEtran}
\IEEEoverridecommandlockouts
\usepackage{multirow}
\usepackage{array}
\usepackage{cite}
\usepackage{amsmath,amssymb,amsfonts}
\usepackage{algorithmic}
\usepackage{graphicx}
\usepackage{textcomp}
\usepackage{xcolor}
\usepackage{fancyhdr}
\usepackage{lipsum}

\def\BibTeX{{\rm B\kern-.05em{\sc i\kern-.025em b}\kern-.08em
    T\kern-.1667em\lower.7ex\hbox{E}\kern-.125emX}}

\begin{document}
\title{Non-Interchangeability between Heart Rate Variability and Pulse Rate Variability During Supine-to-Stand Tests \\
\author{$\text{Runwei Lin}^{*}$, $\text{Frank Halfwerk}^{\dagger,\sharp}$, $\text{Gozewijn Dirk Laverman}^{*,\ddagger}$, $\text{Dirk Donker}^{\S}$, $\text{Ying Wang}^{*}$}
\thanks{
This work has been submitted to the IEEE for possible publication. Copyright may be transferred without notice, after which this version may no longer be accessible.

*Department of Biomedical Signals and Systems, University of Twente, The Netherlands. (e-mail: r.lin@utwente.nl; ying.wang@utwente.nl). $^{\dagger}$Group of Engineering Organ Support Technologies, University of Twente, The Netherlands, $^{\sharp}$Department of Thorax Centrum Twente, Medisch Spectrum Twente, The Netherlands. $^{\ddagger}$Department of Internal Medicine, Ziekenhuisgroep Twente, The Netherlands.
$^{\S}$Group of Cardiovascular and Respiratory Physiology, University of Twente, The Netherlands.}
}

\maketitle

\begin{abstract}
Heart rate variability (HRV) is widely recognized as a valuable biomarker for assessing autonomic cardiac regulation. Pulse rate variability (PRV) is a common surrogate of HRV given the wide usability of PPG in commercially available devices. However, there is no clear conclusion on whether PRV can replace HRV given their different physiological mechanisms. This study evaluates the interchangeability of young adults’ HRV and PRV during supine-to-stand (STS) tests which are known as common posture transitions in daily life monitoring. Fifteen features from time, frequency and nonlinear domains were extracted from both electrocardiography and PPG signals. Paired t-tests and Wilcoxon signed-rank tests examined the difference between the extracted HRV and PRV features during supine, transition and standing phases separately. One feature showed significant difference in the supine phase, and this discrepancy increased to four in the transition and standing phases. These findings suggested that PRV is different from HRV in the STS tests, despite the fact that both metrics can reflect the sympathetic activation triggered by the posture changes.
\end{abstract}


\section{Introduction}
Heart rate variability (HRV) is one of the most widely used biometrics for autonomic tone evaluation \cite{rajendra2006heart}. It can be quantified in the time, frequency, and nonlinear domain. HRV has been linked to diverse types of physiological and pathological factors, including physical and mental stress, respiratory dynamics, and cardiovascular disorders \cite{kim2018stress,aysin2006effect,rajendra2006heart}.  Accordingly, the ambulatory HRV monitoring holds substantial promise as a non-invasive tool for cardiovascular health monitoring. 

HRV analysis usually requires electrocardiography (ECG) measurement with wet electrode patches. However, the gel on the electrodes can stimulate skin, causing uncomfortableness and is therefore cumbersome in daily life monitoring. In recent years, photoplethysmography (PPG) has been widely applied in daily life monitoring given its high usability. PPG measures the relative change in arterial blood volume from sites such as fingers or wrist \cite{charlton2022wearable}. Pulse rate variability (PRV) derived from PPG has also been applied in both researches and daily monitoring applications as a surrogate of HRV since the pulsatile component of the PPG signal fluctuates with heartbeats.

Despite of its wide application, there is no clear conclusion on whether PRV can replace HRV. A recent study reviewed 38 articles examining the relationship between HRV and PRV \cite{mejia2020pulse}. Eighteen studies concluded that PRV is not identical to HRV. One major factor contributing to the difference between HRV and PRV is pulse arrival time (PAT), which refers to the time it takes for the pulse wave to travel from the R peak in the ECG to the point of pulse detection. Blood pressure (BP) change has been identified as a crucial factor that can affect PAT, and therefore the PRV-HRV difference \cite{gil2010photoplethysmography}.  

Supine-to-stand (STS) test is a widely used method to induce BP change. It involves transitioning from a supine to a standing position, triggering sympathetic activation and typically reducing HRV. Given its relevance to real-world posture changes and BP variations, analyzing HRV-PRV interchangeability during STS can enhance our understanding of autonomic cardiac regulation and inform PRV applications in daily life monitoring. This study thus investigates the interchangeability between HRV and PRV during STS tests.

\section{Material and methods}
\subsection{Data pre-processing}
The dataset used in this study was adopted from \cite{mol2020pulse} and was approved by the ethics committee of Radboud University (ECS17022 and REC18021). The young group that performed the STS tests was included in the study (Age 22-45 years, median 25). In each session, the participants lied down in a supine position for five minutes firstly and then stood up and kept an upright position for three minutes. PPG and ECG signal were recorded under a sampling rate of 1000Hz and 300Hz respectively.

\begin{figure}[htbp]
    \centering
    \includegraphics[width=0.75\linewidth]{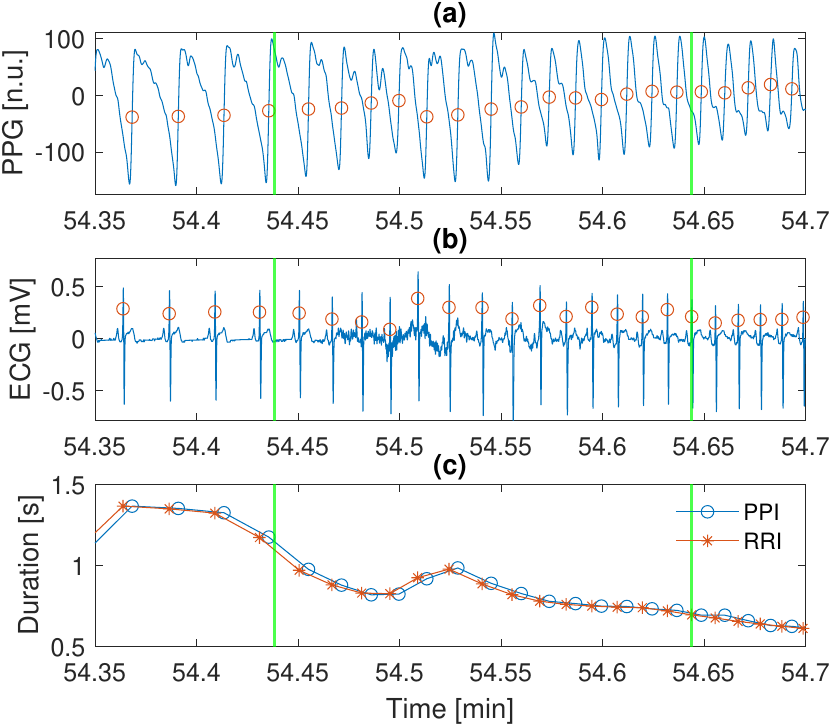}
    \caption{A representative example of ECG and PPG measurements, along with the PPI and RRI durations.  Green lines indicate the end of supine and the end of subjects standing up. (a) PPG signal, red dots represent the fiducial points. (b) ECG signal, red dots donates detected R peaks. (c) The change of PPI and RRI duration.
   }
    \label{signal_eg}
\end{figure}

Fig. \ref{overview} illustrates data pre-processing process. We applied the MSPTDfast algorithms to identify the fiducial points from PPG signals given its superior performance \cite{charlton2022detecting,charlton2024msptdfast}. The midpoints between the diastole and systole peaks were used as the fiducial points of the due to its robustness to noise \cite{charlton2022detecting}. The distance between each two midpoints was defined as pulse peak intervals (PPI). Pan-Thompkins method was utilized to identify the R peaks from Lead II ECG \cite{pan1985real}, and the distance between each two R peaks was defined as R-R intervals (RRI). The PPI and RRI sequences were further smoothed by removing outliers, using 'RRfilter' in the toolbox introduced in \cite{vollmer2019hrvtool}. Fig. \ref{signal_eg} exhibits a representative example of measured ECG and PPG and the derived PPI and RRI duration. 

Since PPG signal is usually sensitive to motion artifacts, we adopted the strategy from \cite{mejia2021differential} to select high quality PPG data.  
A signal quality index (SQI) from \cite{li2012dynamic} was computed for each detected PPI. A K-means clustering method divides all SQIs into two groups: high SQI group and low SQI group. The percentage of high quality points can be then calculated as follows: 
\begin{equation}
\label{gd}
    R_{good} = 100\%\cdot\frac{N_{good}}{N}
\end{equation}
where $N_{good}$ is the number of PPIs in high SQI groups, $N$ is total number of PPIs. If the ‘good ratio’ $R_{good}$ is above 80\% \cite{mejia2021differential} or the mean SQI above 0.8, which is empirically determined, then the data were considered high quality and used for the following analysis. In total, 30 STS sessions were analyzed. 

\subsection{Interchangeability analysis}
\begin{figure}[h]
    \centering
\includegraphics[width=0.85\linewidth]{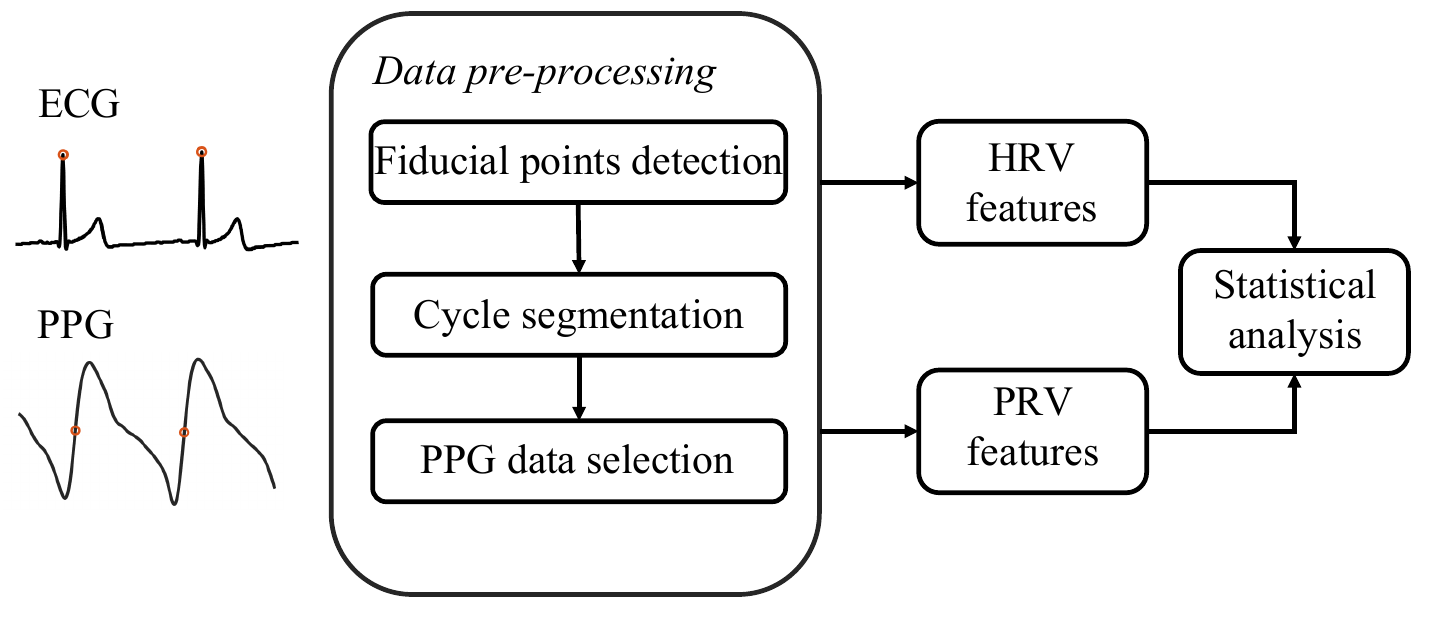}
    \caption{Overview of the ECG and PPG signal processing and analysis.}
    \label{overview}
\end{figure}

\begin{figure}[htbp]
    \centering
    \includegraphics[width=0.6\linewidth]{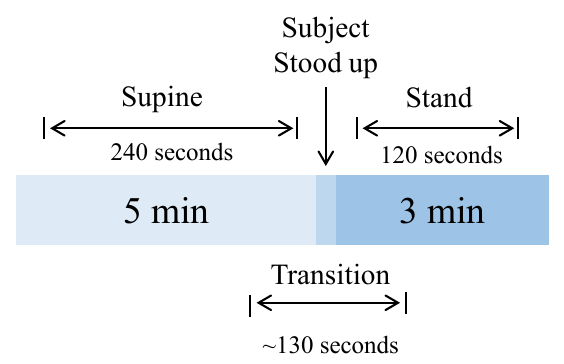}
    \caption{Illustration of the experiment protocol and the windows selected for HRV and PRV analysis. }
    \label{protocol}
\end{figure}

Fig. \ref{overview} shows an overview of the ECG and PPG analysis. 
After data pre-processing, data from each STS session were divided into three phases: supine, transition, and standing, as illustrated in Fig. \ref{protocol}. 
To minimize interference caused by postural changes inter- and intra-experimental sessions, we analyzed the supine phase data excluding its first and the end 30 second data, the transition phase data including each one-minute data before and after the subject stood up, and the standing phase data from the 30 seconds after subject stood up until 30 seconds before the end of the whole session. 
 
HRV features were subsequently calculated within three phases. A total of fifteen commonly used features spanning time, frequency, and nonlinear domains were included, as summarized in Table \ref{parameters}. Time and nonlinear features was implemented using the toolbox from \cite{vollmer2019hrvtool}. Frequency domain features, i.e., LF, HF, nLF, nHF and LF/HF were calculated by fast Fourier transform after interpolating the RR sequences to 4 Hz by piecewise cubic splines, as suggested in \cite{mejia2022spectral}. Both approximate entropy and sample entropy were calculated using embedding dimension of two, and a tolerance of 20\% of the data standard deviation, in line with previous studies \cite{RN13}. 
\begin{table}[htbp]
	\caption{Summary of used HRV and PRV features}
	\label{parameters}
	\centering
	\begin{tabular}{ll} 
		\hline 
		\textbf{Feature [unit]} & \textbf{Definition} \\ 
		\hline
		AHR [bpm] & Average heart rate \\ 
		RMSSD [ms] & Root mean square of the differences between successive\\& RR intervals \\ 
		SDNN [ms] & Standard deviation of all RR intervals \\ 
		SDSD [ms] & Standard deviation of the differences between successive \\& RR intervals \\ 
		pNN50 [\%] & Percentage of successive RR intervals where the difference \\&exceeds 50 milliseconds \\ 
		LF [$\text{ms}^2$] & Low frequency band power \\ 
		HF [$\text{ms}^2$] & High frequency band power \\ 
		nLF & Normalized low frequency band power \\ 
		nHF & Normalized high frequency band power \\ 
		LF/HF & Ratio between low and high frequency band power \\ 
		SD1 [ms] & Short-term variability derived from the Poincaré plot \\  
		SD2 [ms] & Long-term variability derived from the Poincaré plot \\ 
		SD2/SD1 & Ratio between SD2 and SD1 \\ 
		ApEn & Approximate entropy, which quantifies the regularity of\\& RR intervals \\ 
		SampEn & Sample entropy: a refined version of approximate entropy \\ 
		\hline
	\end{tabular}
\end{table}

Paired T-tests were used to assess the difference between HRV and PRV features that followed a normal distribution, which was determined by Shapiro-Wilk tests at a significance level of 0.05. For features that did not follow a normal distribution, the non-parametric Wilcoxon signed-rank tests were employed to evaluate their difference. Since 15 tests were performed for each feature in each phase, the significance level was adjusted to 0.05/15 = 0.0033 after Bonferroni correction to avoid type I error.

\section{Results}
\begin{figure*}[htbp]
    \centering   \includegraphics[width=0.95\linewidth]{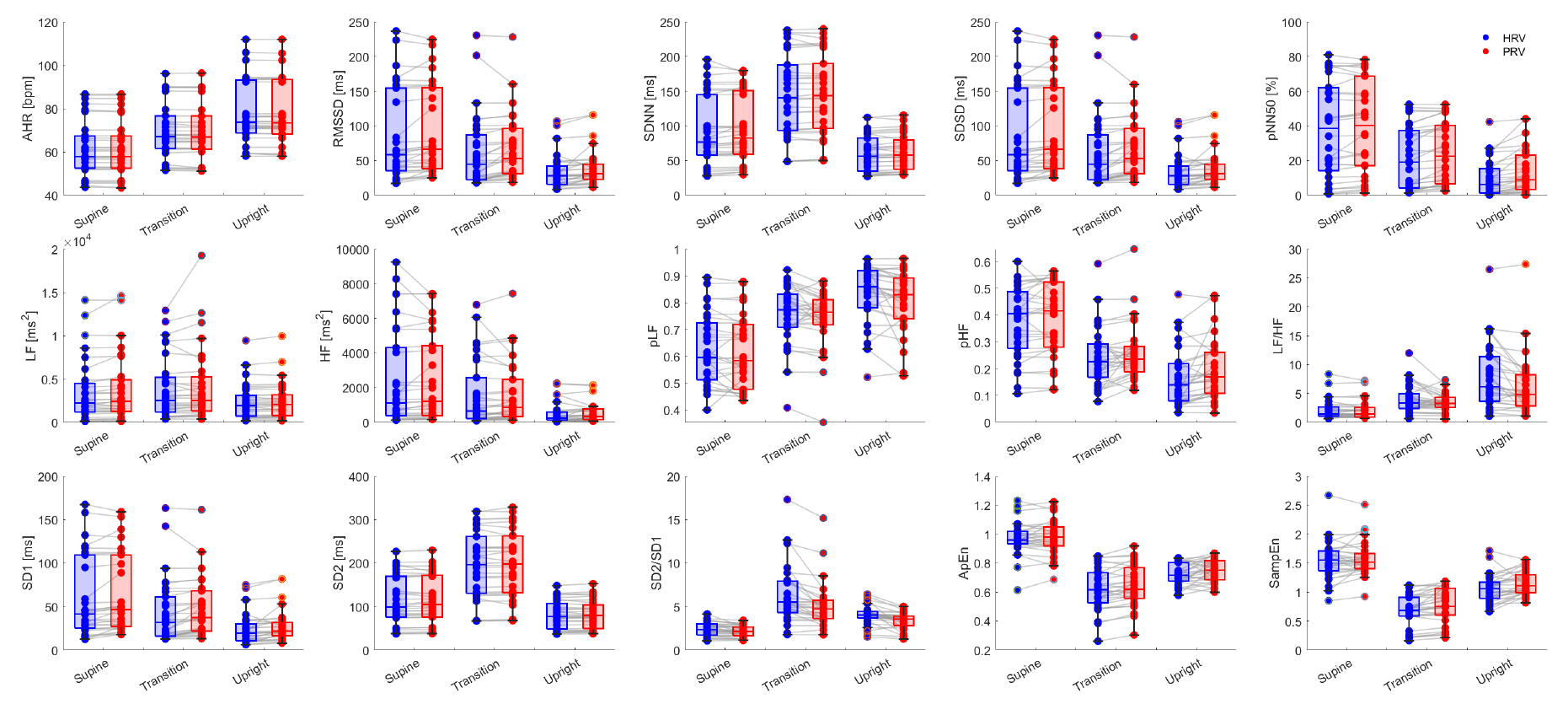}
    \caption{Comparison of the distribution of all calculated HRV and PRV features across supine, transition, and standing phases (whisker length = 1.5).}
    \label{young}
\end{figure*}

Fig. \ref{young} shows the boxplot comparison between HRV and PRV features across all three phases. The features that positively relates to sympathetic activity level, such as AHR, pLF and LF/HF, have a clearly increasing trend. Features that are negatively related to the sympathetic arousal, such as RMSSD, SDSD, pNN50, HF, pHF and SD1, significantly decreases during the test.

Table \ref{pvalue} summarizes the p-values from the statistical analysis, with values below 0.0033 marked by an asterisk. During the supine phase, one features (SD2/SD1) exhibited significant differences. In both transition and the standing phases, four features have significant differences. 

SD2/SD1 exhibited significant differences across all three phases, suggesting it may be influenced by additional peripheral factors. HF, pNN50, ApEn, and SampEn showed no significant differences in the supine phase but became significant in later phases, indicating that these PRV features are largely affected by pulse wave propagation. Overall, nonlinear indices demonstrated more significant differences compared to time and frequency domain indices.
\begin{table}[htbp]
\caption{p-values of the statistical tests. The significances were marked with *}
\label{pvalue}
\centering
\begin{tabular}{lccc}
\hline 
 & Supine & Transition & Standing \\
\hline
AHR &0.1353& 0.0630 & 0.0552\\
RMSSD &0.0086&0.0180 &0.0067 \\
SDNN &0.1535&0.0316 &0.0057\\
SDSD &0.0085&0.0186 &0.0065 \\
pNN50 & 0.1032&$0.00002^{*}$ &$0.0006^{*}$\\
LF & 0.0346&0.2761 & 0.1229 \\
HF & 0.8273& 0.9719 & 0.2208\\
nLF & 0.1846& 0.1058& 0.0151 \\
nHF  & 0.1846&0.1058 & 0.0133\\
LF/HF & 0.1879&0.0256& $0.0015^{*}$\\
SD1 & 0.0085&0.0186 & 0.0065 \\ 
SD2& 0.0125& 0.0428 & 0.1156\\
SD2/SD1 &$0.0003^{*}$&$0.0003^{*}$ & $0.00003^{*}$\\
ApEn &0.6583& $0.0028^{*}$ & 0.0407 \\
SampEn &0.3034&$0.0001^{*}$ &$0.0015^{*}$ \\
\hline
\end{tabular}
\end{table}
\section{Discussion and conclusions}
This study explores the interchangeability of HRV and PRV features during STS tests in young adults. Parametric and non-parametric methods evaluates the difference between HRV and PRV across different phases. One metrics showed significant difference during supine phase. After the subject standing up, this number increased to four in the transition and standing phases. The results suggest that PRV should not be used as a surrogate for HRV especially during the postural transition phases, although both metrics can reflect the autonomic tone change.   

The results of pNN50, LF/HF, ApEn and SampEn changed from non-significant to significant from supine to transition phase. This discrepancy may be attributed to the change of sympathetic representation in PRV measurements due to pulse wave propagation and the peripheral vascular effects during STS tests. Overall, nonlinear indices exhibited more significant differences compared to time and frequency domain features. Interestingly, SD2/SD1 presented significant difference in all three phases. This might imply it reflects the intrinsic difference between HRV and PRV. Further studies are needed to explore the physiological implications of these patterns, particularly in dynamic postural changes.

Fig. \ref{young} exhibits the HRV and PRV change during the three phases. It can be observed that both PRV and HRV exhibited consecutive increases in AHR, nLF, and LF/HF from the supine to the upright position. Conversely, RMSSD, SDSD, pNN50, LF, HF, nHF, and SD1 progressively decreased during this transition. These observations demonstrate that both HRV and PRV effectively reflect sympathetic activation during postural change.  An interesting observation is that for most features that presented a decreasing trend in this study, such as RMSSD, SDSD, pNN50, HF, and SD1, their variability—reflected by the interquartile range—also decreases. Similarly, features with an increasing trend, such as AHR and LF/HF, display a corresponding increase in variability. These trends might be explained by reduced inter-subject variability in parasympathetic activity and increased inter-subject variability in sympathetic activity during sympathetic activation. 

This study has certain limitations. The HRV and PRV were calculated using window lengths of two to four minutes. Although duration less than five minutes are relatively short for traditional HRV analysis \cite{RN54}, they were chosen to balance the trade-off between better dynamic capture and reliable analysis. Previous studies also reported the selected lengths (above 90 second) is sufficient for the PRV analysis \cite{mejia2023duration}. In conclusion, our results suggest that PRV should not be used as a surrogate for HRV, even though both metrics reflect changes in autonomic tone during the postural change in young participants. 

Results of our study support the previous findings \cite{RN19}. It should be cautious to use PRV to replace HRV in non-stationary applications. Further studies are therefore necessary to investigate the performance of PRV in traditionally HRV based tasks, such as autonomic function assessment and stress evaluation. However, it is also important to acknowledge that PRV may also contain additional information related to hemodynamics and cardiopulmonary coupling, which is not captured by HRV. This unique aspect of PRV suggests its potential to enhance daily cardiovascular health monitoring, including the management of conditions like cardiovascular and metabolic diseases such as diabetes mellitus. Future studies could focus on exploring PRV differences between healthy individuals and patients with specific conditions to uncover its broader applications in daily health monitoring.
\section*{Acknowledgment}
The authors would like to acknowledge prof. dr. Richard van Wezel and dr. Arjen Mol from Donders Institute for Brain, Cognition and Behaviour for providing the dataset. This study is in part by China Scholarship Council (No. 202308310039).
\bibliographystyle{IEEEtran}
\bibliography{refs}
\end{document}